\documentstyle[preprint,aps,epsf,floats]{revtex} 

\newcommand{\nn}{\nonumber} 

\begin{document}
\setlength\baselineskip{20pt}

\preprint{\tighten \vbox{\hbox{CALT-68-2173}
		\hbox{hep-ph/yymmxxx} }}

\title{ A momentum subtraction scheme for two--nucleon \\[-5pt]
  effective field theory   }

\author{Thomas Mehen and Iain W.\ Stewart \\[4pt]}
\address{\tighten California Institute of Technology, Pasadena, CA 91125 }

\maketitle

{\tighten
\begin{abstract} 

We introduce a momentum subtraction scheme which obeys the power counting of
Kaplan, Savage, and Wise (KSW), developed for systems with large scattering
lengths, $a$. Unlike the power divergence subtraction scheme, coupling constants
in this scheme obey the KSW scaling for all $\mu_R > 1/a$.  We comment on the
low-energy theorems derived by Cohen and Hansen.  We conclude that there is no
obstruction to using perturbative pions for momenta $p>m_\pi$.

\end{abstract}
}
\vspace{0.7in}

\newpage

Effective field theory is a useful method for describing  two-nucleon systems
\cite{orefs}.  Recently, Kaplan, Savage, and Wise (KSW) \cite{ksw} devised a power
counting that accounts for the effect of large scattering lengths.  With this power
counting the dimension six four-nucleon operators are treated non-perturbatively,
while pion exchange as well as higher dimension operators are treated
perturbatively.  A key feature of the KSW approach is the use of the power
divergence subtraction (PDS) renormalization scheme.  PDS gives the coefficients
of contact interactions power law dependence on the renormalization scale, which
makes the KSW power counting manifest.  For $\mu_R \gtrsim 300\,{\rm MeV}$, the
power counting in PDS is no longer manifest \cite{ksw}, and KSW conclude that
application of the theory is restricted to momenta less than $300\,{\rm MeV}$.  

In Ref.~\cite{Bira}, it is emphasized that the power counting for large scattering
lengths should be scheme independent.  Here we introduce a momentum
subtraction scheme that is compatible with the KSW power counting.  A similar
scheme is applied to the theory without pions in Ref.~\cite{Gegelia1}.  This
scheme will be referred to as the OS scheme, since in a relativistic field theory it
would be called an off-shell momentum subtraction scheme.  One attractive
feature of the OS scheme is that the KSW power counting is manifest for all
$\mu_R>1/a$, where $a$ is the scattering length.  Since the breakdown of the
power counting is an artifact of the PDS scheme, the range of the effective field
theory with perturbative pions may not be limited to $300\,{\rm MeV}$.  

Another attractive feature of the OS scheme is that the amplitudes are manifestly
$\mu_R$ independent at each order in the expansion.  In PDS, any fixed order
calculation has residual $\mu_R$ dependence which is cancelled by higher order
terms.  The rather strong $\mu_R$ dependence exhibited by next-to-leading order
calculations of the phase shifts led authors \cite{Gegelia2,sf2} to conclude that
PDS results have to be fine tuned to fit the data.  In fact, it is possible to fix this
problem with PDS by modifying the treatment of $C_0(\mu_R)$.  

We begin our discussion by recalling the Lagrangian with pions and nucleons
\cite{ksw}
\begin{eqnarray} 
{\cal L}_\pi &=& \frac{f^2}{8} {\rm Tr}\,( \partial^\mu\Sigma\: \partial_\mu 
\Sigma^\dagger )+\frac{f^2\omega}{4}\, {\rm Tr} (m_q \Sigma+m_q \Sigma^\dagger) - \frac{ig_A}2\, N^\dagger \sigma_i (\xi\partial_i\xi^\dagger -
\xi^\dagger\partial_i\xi) N  \nn\\ 
&+& N^\dagger \bigg( i D_0+\frac{\vec D^2}{2M} \bigg) N -  {C_0^{(s)}} ( N^T P^{(s)}_i N)^\dagger ( N^T P^{(s)}_i N) \label{Lpi} \\ 
&-&   {C_2^{(s)}\over 8} \left[ ( N^T P^{(s)}_i N)^\dagger ( N^T
P^{(s)}_i \:\tensor{\nabla}^{\,2} N) + h.c. \right] -{D_2^{(s)}} \omega {\rm
Tr}(m^\xi ) ( N^T P^{(s)}_i N)^\dagger ( N^T P^{(s)}_i N) + \ldots \nn . 
\end{eqnarray} 
Here $g_A=1.25$ is the nucleon axial-vector coupling, $f=130\, {\rm MeV}$ is the
pion decay constant, $m^\xi=\frac12(\xi m_q \xi + \xi^\dagger m_q \xi^\dagger)$,
where $m_q={\rm diag}(m_u,m_d)$ is the quark mass matrix, and $m_\pi^2 =
w(m_u+m_d)$.  The matrices $P_i^{(s)}$ project onto states of definite spin and
isospin, and the superscript $s$ denotes the partial wave amplitude mediated by
the operator.  This paper will be concerned only with S-wave scattering, so
$s={}^1\!S_0, {}^3\!S_1$. The ellipsis in Eq.~(\ref{Lpi}) denotes terms that are 
higher order in the expansion.

 The four nucleon couplings $C_0$, $C_2$, and $D_2$  in Eq.~(\ref{Lpi}) are bare
parameters. In general, a bare coupling, $C^{\rm bare}$, can be separated into a
renormalized coupling, $C(\mu_R)$, and counter\-terms: 
\begin{eqnarray}
C^{\rm bare} &=&C^{\rm finite}-\delta^{\rm uv}C\,, \qquad\quad C^{\rm finite}
 =C(\mu_R) -\sum_{n=0}^\infty\delta^n C(\mu_R) \,. 
\label{ctexpn} 
\end{eqnarray} 
The counterterms have been divided into two classes. The first, which have the
superscript uv, contain all ultraviolet divergences and are $\mu_R$ independent. 
Defining a renormalization scheme amounts to making a choice for the finite
counterterms, $\delta^nC(\mu_R)$.  The superscript $n$ indicates that $\delta^nC$
is included at tree level for a graph with $n$ loops.  For example, at two loops, we
have two loop diagrams with renormalized couplings at the vertices, one loop
diagrams with a single $\delta^1 C$ counterterm, and a tree level diagram with
$\delta^2 C$.    Let $Q$ denote a typical momentum characterizing the process
under consideration.  For nucleon-nucleon scattering we take $p\sim Q$ and
$m_\pi\sim Q$, where $p$ is the center-of-mass momentum. The theory is an
expansion in $Q/\Lambda$ where $\Lambda$ is the range of the effective field
theory. Taking $\mu_R \sim Q$, vertices with $C_0(\mu_R)$ scale as $1/Q$, while
vertices with $C_2 p^2$ or $D_2 m_\pi^2$ scale as $Q^0$.  A typical loop gives
one power of $Q$, so $C_0(\mu_R)$ vertices are included to all orders.  This sums
all corrections that scale as $(Qa)^n$ \cite{ksw}.  Note that since the pion has
been included explicitly in the Lagrangian we expect that the scale of short
distance physics, $\Lambda$, should not be set by $m_\pi$, but by higher mass
resonances which have not been included in the theory.  

PDS is one scheme in which the KSW power counting is manifest.  In PDS, we first
let $d=4-2\epsilon$ and define the counterterms $\delta^{\rm uv}C$ to subtract
$1/\epsilon$ poles.  As in the $\overline{\rm MS}$ scheme, the dimensional
regularization parameter $\mu$ is set to $\mu_R$.  Next one takes $d=3$ and
defines the finite counterterms, $\delta^nC(\mu_R)$ to subtract the $1/(d-3)$ poles
in the amplitude.  Graphs which renormalize a given coupling are those whose
vertices have the right number of derivatives and powers of $m_\pi^2$.  When
calculating $\delta^nC_0(\mu_R)$ and $\delta^nC_2(\mu_R)$, we can take
$m_\pi=0$ since counterterms proportional to $m_\pi^2$ renormalize coefficients
like $D_2(\mu_R)$.  After making these subtractions, the amplitude is continued
back to four dimensions, so $d-3\to 1$.  

In the $^1S_0$ channel, exact expressions for the PDS beta functions can be
obtained \cite{ksw}.  For $C_0(\mu_R)$ we have
\begin{eqnarray} 
  C_0(\mu_R) = \frac{4\pi}{M} \left( \frac1{K-\mu_R} \ - \frac{1}{\Lambda_{NN}} 
  \right ) \,,
\end{eqnarray}
where $\Lambda_{NN} = 8 \pi f^2/(M g_A^2) = 300\,{\rm MeV}$. $K$ is a constant
fixed by the boundary condition, and choosing $C_0(0) = 4\pi a/M$ gives $K = 1/(a
+1/\Lambda_{NN})$.  We see that the scaling for $C_0(\mu_R)$ changes for $\mu_R
\sim 300\, {\rm MeV}$\cite{ksw}.  A simple shift, $C_0(\mu_R)\to C_0(\mu_R) -
g_A^2/2f^2$, results in a coupling that scales as $1/\mu_R$ for all $\mu_R > 1/a$. 
Physically, this shift corresponds to summing the short distance ($m_\pi=0$)
contributions from potential pion exchange to all orders.  For the ${}^3\!S_1$
channel, this summation is not possible because of ultraviolet divergences of the
form $p^2/\epsilon$ and, in fact, there are unknown corrections to the $^3S_1$
beta functions at each order in $Q$.  This is demonstrated in Ref.~\cite{ms1} by
explicit computation of the counterterms.  When the beta function is not exactly
known, the large $\mu_R$ behavior is ambiguous.  For example, the PDS beta
function for $C_0(\mu_R)$ is
\begin{eqnarray}  
\label{b0} 
\beta_0 &=&  \mu_R \frac{\partial C_0(\mu_R)}{\partial \mu_R} = 
     \frac{M\mu_R}{4\pi}\:  \left[ C_0^2(\mu_R)+ 2\frac{g_A^2}{2f^2} C_0(\mu_R) 
     \right] + {\cal O}(Q)  \,.  
\end{eqnarray} 
Two solutions which satisfy this equation to order $Q^0$ are 
\begin{eqnarray}  
\label{gsln}
  C_0(\mu_R) &=& -\,\frac{4\pi}{M} \frac{\Big[ 1-2 a \mu_R-2\mu_g - \sqrt{1+4
    \mu_g^2 -4 \mu_g (1-a\mu_R)} \ \Big]}{2\mu_R (1-a\mu_R -\mu_g)} \,,  \nn \\
  C_0(\mu_R) &=& -\,\frac{g_A^2}{f^2} \frac{1}{1- \Big[1+2/(a \Lambda_{NN} )\Big] 
  \exp{(-2\mu_g)} } \,, 
\end{eqnarray}
where $\mu_g = \mu_R/\Lambda_{NN}$ and we have chosen $C_0(0)=4\pi
a/M$.  The first solution is obtained by computing the counterterms $\delta^n
C_0(\mu_R)$ to order $Q^0$ and summing them.  This solution falls as $1/\mu_R$
for all $\mu_R > 1/a$, and is numerically close to the $g_A\to 0$ solution.  The
second solution is obtained by truncating and solving Eq.~(\ref{b0}).  This solution
approaches a constant as $\mu_R\to \infty$.  The two solutions both solve the
beta function to order $Q^0$ but have very different large $\mu_R$ behavior.  In
the OS scheme, there is no ambiguity since at a given order in $Q$ the running of 
all the coupling constants that enter is known exactly.  

In the OS momentum subtraction scheme, the renormalized couplings are defined
by relating them to the amplitude evaluated at the unphysical momentum
$p=i\mu_R$.  We start by dividing up the full amplitude as $A = \sum_{m=0}^\infty
A^{(m-1)}$.  Here $A^{(m-1)}$ contains the Feynman diagrams that scale as
$Q^{m-1}$.  As in PDS, we can take $m_\pi\to 0$ when computing the
counterterms for $C_0(\mu_R)$ and $C_2(\mu_R)$.  $\delta^{\rm uv}C_{2m}$ is first
defined to subtract all four dimensional $1/\epsilon$ poles.  
 \begin{figure}[t!]  
  \centerline{\epsfysize=8.2truecm \epsfxsize=16.0truecm \epsfbox{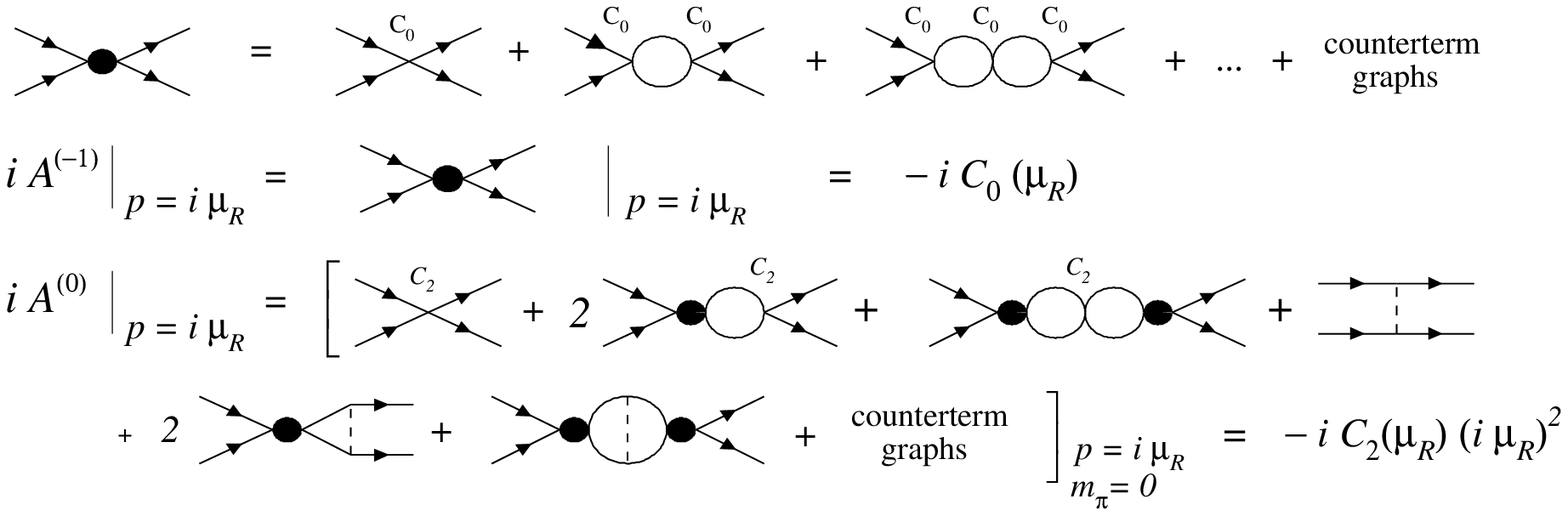}} 
{\tighten
 \caption[1]{Renormalization conditions for $C_0(\mu_R)$ and $C_2(\mu_R)$ in 
the OS scheme.  $i\,A^{(-1)}$ is the four point function with $C_0(\mu_R)$
 and $\delta^n C_0(\mu_R)$ vertices, evaluated between incoming and 
 outgoing  ${}^1S_0\mbox{ or }{}^3S_1$ states.  The amplitude $A^{(0)}$ 
 contains graphs with one $C_2$ or one potential pion dressed with $C_0$ 
 bubbles.}  \label{fig_C0} }  
\end{figure}  
The definition for the renormalized coupling is then 
\begin{eqnarray} 
  & & i  A^{(m-1)}  \bigg|_{\mbox{\scriptsize$\begin{array}{c} \ p=i\mu_R \\[-4pt] 
   \!\!\!\! m_\pi=0  \end{array}$}} = -i  C_{2m}(\mu_R)\, (i\mu_R)^{2m}\,.  \label{rc2m}
\end{eqnarray} 
This condition is to be imposed order by order in the loop expansion so that
the graphs at $n$ loops determine $\delta^nC_0(\mu_R)$.  For instance, the
couplings $C_0(\mu_R)$ and $C_2(\mu_R)$ are defined by the renormalization
condition in Fig.~\ref{fig_C0}.  Summing the counterterms, we find 
\begin{eqnarray} \label{OSc0c2} 
  C_0(\mu_R) &=& {C_0^{\rm finite} \over 1 -\mu_R
         \frac{C_0^{\rm finite} M}{(4\pi)} } \,, \qquad\quad 
  C_2(\mu_R) = {C_2^{\rm finite} - {g_A^2}/({2f^2\mu_R^2}) \over 
         \left[ 1 - \mu_R \frac{C_0^{\rm finite} M}{4\pi}   \right]^2 }\,.  
\end{eqnarray}
Although it may seem that the piece of $C_2(\mu_R)$ that goes as
$1/\mu_R^4$ will spoil the power counting for low momentum, in fact, the
$1/\mu_R^2$ part dominates entirely for $\mu_R\gtrsim 1/a$, since $C_0^{\rm
finite} \sim a$, $C_2^{\rm finite}\sim a^2$.

In the OS scheme, $D_2(\mu_R)$ will be calculated as follows.  First
$m_\pi^2/\epsilon$ poles are subtracted.  The renormalized coupling is then 
defined by 
\begin{eqnarray}
   &&  i\,A(D_2) \Big|_{p=i\mu_R,} =  -i\, D_2(\mu_R)\, m_\pi^2 \,,
\end{eqnarray}
where $A(D_2)$ contains terms in the amplitude that are proportional to $m_\pi^2$,
as well as analytic in $m_\pi^2$.   Thus, at $Q^0$ graphs with a single
$D_2(\mu_R)$ or potential pion and any number of $C_0(\mu_R)$ vertices
contribute to $A(D_2)$.  Note that we only keep terms that are analytic in
$m_\pi^2$ because it seems unnatural to put long-distance nonanalytic
contributions that come from pion exchange into the definition of the short
distance coupling $D_2(\mu_R)$ \cite{sf2}.  For example,  one potential pion
exchange gives a $(m_\pi^2/p^2) \ln(1+4p^2/m_\pi^2)$ term.  Putting this into
$D_2(\mu_R)$ would give it both a branch cut at $\mu_R=m_\pi/2$ as well as
explicit dependence on the scale $m_\pi$, therefore this term is not included in
$A(D_2)$.  Solving for the counterterms and summing we find 
\begin{eqnarray} 
 {D_2(\mu_R)\over C_0(\mu_R)^2} &=& { D_2^{\rm finite} \over 
   (C_0^{\rm finite})^2}  +\frac{M}{8\pi}  \bigg(\frac{M g_A^2}{8\pi f^2} \bigg) \left[
   \ln\bigg({\mu_R^2 \over \mu_0^2}\bigg) -1 \right] = \frac{M}{8\pi} 
  \bigg(\frac{M g_A^2}  {8\pi f^2} \bigg)  \ln\bigg({\mu_R^2  \over \tilde \mu^2}
  \bigg)  \,, \nn\\[5pt]
 && \quad \mbox{  where    } \tilde\mu^2 = \mu_0^2 \, \exp{\bigg(1 - \frac{64\pi^2 
  f^2  D_2^{\rm finite}}{M^2  g_A^2(C_0^{\rm finite})^2} \bigg)} \,.  \label{D2} 
\end{eqnarray} 
With $m_\pi\sim Q\sim \mu_R$, $D_2(\mu_R)m_\pi^2 \sim Q^0$. The scale 
$\tilde \mu$ must be determined by fitting to data.  The PDS solution for 
$D_2(\mu_R)$ does not have the $-1$ in square brackets.

We now compare the amplitudes in the PDS and OS schemes.  At order $Q^0$, the
amplitudes in the ${{}^1\!S_0}$ and ${{}^3\!S_1}$ channels have the same
functional form.  The amplitudes in PDS are calculated to this order in
Ref.~\cite{ksw}.  We find
\begin{eqnarray}  
  A &=& A^{(-1)} + A^{(0,a)} + A^{(0,b)} + {\cal O}(Q^2) \,, \nn \\[5pt]
  {A^{(-1)}} &=& -\frac{4\pi}{M}\: \frac{1} {\frac{4\pi}{M C_0(\mu_R)} +
	\mu_R + ip } \,, \label{Amp1} \\[10pt]  
 \frac{A^{(0,a)}}{\Big[{A^{(-1)}}\Big]^2} &=& 
  \frac{ g_A^2 m_\pi^2 }{2 f^2} \bigg( \frac{M}{4\pi} \bigg)^2 \left\{ 
   \frac12 \ln{\bigg({ \mu_R^2\over m_\pi^2} \bigg)} - \bigg(\frac{4\pi}
  {MC_0(\mu_R)} + \mu_R\bigg) \frac{1}{p} \tan^{-1}\bigg(\frac{2p}{m_\pi}\bigg)
    \right.  \\[5pt]
&&\left.\  + \bigg[ \bigg(\frac{4\pi}{MC_0(\mu_R)} 
  +\mu_R\bigg)^2-p^2 \bigg] \frac1{4p^2} \ln{\bigg(1+ \frac{4p^2}{m_\pi^2} 
  \bigg)} \right\} - \frac{D_2(\mu_R)\: m_\pi^2 }{C_0(\mu_R)^2}  \,, \nn \\[5pt]
\mbox{PDS}\quad \frac{A^{(0,b)}}{\Big[{A^{(-1)}}\Big]^2} &=& 
  -\frac{ C_2(\mu_R)\: p^2} {C_0(\mu_R)^2}  -\frac{g_A^2}
  {2 f^2}  \frac1{C_0(\mu_R)^2} + \frac12 \: \frac{ g_A^2 m_\pi^2 }{2 f^2} \bigg( 
  \frac{M}{4\pi} \bigg)^2 \,, \label{pdsAb} \\ 
\mbox{OS}\quad \frac{A^{(0,b)}}{\Big[{A^{(-1)}}\Big]^2} &=& 
  -\frac{ C_2(\mu_R)\: p^2} {C_0(\mu_R)^2}  -\frac{g_A^2}
  {2 f^2} \bigg(1+\frac{p^2}{\mu_R^2}\bigg) \bigg( \frac1{C_0(\mu_R)} + 
  \frac{M\mu_R}{4\pi} \bigg)^2  \label{Amp4} \,.
\end{eqnarray}
Note that the last term in Eq.~(\ref{pdsAb}) has a factor of $1/2$ instead of a $1$
as in \cite{ksw} since we have made a different finite subtraction.  The terms in
braces are long distance pion contributions, and are the same in the PDS and OS
schemes.  By substituting Eqs.~(\ref{OSc0c2}) and (\ref{D2}) into the OS amplitude
one can verify that it is $\mu_R$ independent. In contrast, in PDS the amplitude is
only $\mu_R$ independent to order $Q^0$.  

To obtain a good fit to the scattering data at low momenta, two 
constraints must be approximately satisfied:
\begin{eqnarray}  \label{constr}
  \lim_{p \to 0} \ \: p\cot\delta(p) = -\frac{1}{a} \,, \qquad \qquad
  \left.  \frac{A^{(0)}}{[A^{(-1)}]^2} \ \right|_{-ip = \gamma} = 0\, ,
\end{eqnarray}
where $\gamma = \frac{4 \pi}{M C_0(\mu_R)} + \mu_R$.  The second constraint
ensures that the next-to-leading order amplitude does not have a spurious double
pole.   In the PDS scheme, $A^{(0)}$ is not $\mu_R$ independent, so the extracted
parameters can not simultaneously satisfy the renormalization group equations
and give a good fit.  This is the origin of the large dependence on $\mu_R$
observed in Ref.~\cite{Gegelia2}.  In PDS the second constraint gives
\begin{eqnarray}  \label{Mm2PDS}
0&\simeq&{-D_2(\mu_R)  \over C_0(\mu_R)^2} +\frac{g_A^2 }{2 f^2}
  \bigg(\frac{M}{4\pi}\bigg)^2 \bigg[\, \frac12 + \frac12 \ln{\bigg({\mu_R^2 
  \over m_\pi^2}\bigg)} -\frac2{m_\pi}  \bigg(\frac{4\pi}{M C_0(\mu_R)}+ 
  \mu_R \bigg)  \\ 
&& \qquad\qquad\qquad\qquad\qquad\ \ +\frac1{m_\pi^2} \bigg(\frac{4\pi}{M 
   C_0(\mu_R)}+ \mu_R \bigg)^2 - \frac1{m_\pi^2} 
   \bigg(\frac{4\pi}{MC_0(\mu_R)}\bigg)^2 \ \bigg]  \nn \,,
\end{eqnarray}
while in OS we find
\begin{eqnarray}  \label{Mm2OS}
0&\simeq&{-D_2(\mu_R) \over C_0(\mu_R)^2} +\frac{g_A^2}{2 f^2}
 \bigg(\frac{M}{4\pi}\bigg)^2 \bigg[\ \frac12 \ln{\bigg({\mu_R^2 
 \over m_\pi^2}\bigg)} -\frac2{m_\pi} \bigg(\frac{4\pi}{M C_0(\mu_R)}+ 
 \mu_R \bigg) \bigg] \,,
\end{eqnarray}
where terms of order $\gamma^2/m_\pi^2$ have been dropped.   With the
convention for $D_2(\mu_R)$ in Ref.~\cite{ksw} the first term in square brackets in
Eq.~(\ref{Mm2PDS}) is a $1$, so they find that $D_2(m_\pi)$ is close to zero.  The
last term in Eq.~(\ref{Mm2PDS}) induces the large $\mu_R$ dependence.  On the
other hand Eq.~(\ref{Mm2OS}) is $\mu_R$ independent.  

\begin{figure}[!t]  
  \centerline{\epsfxsize=8.0truecm \epsfbox{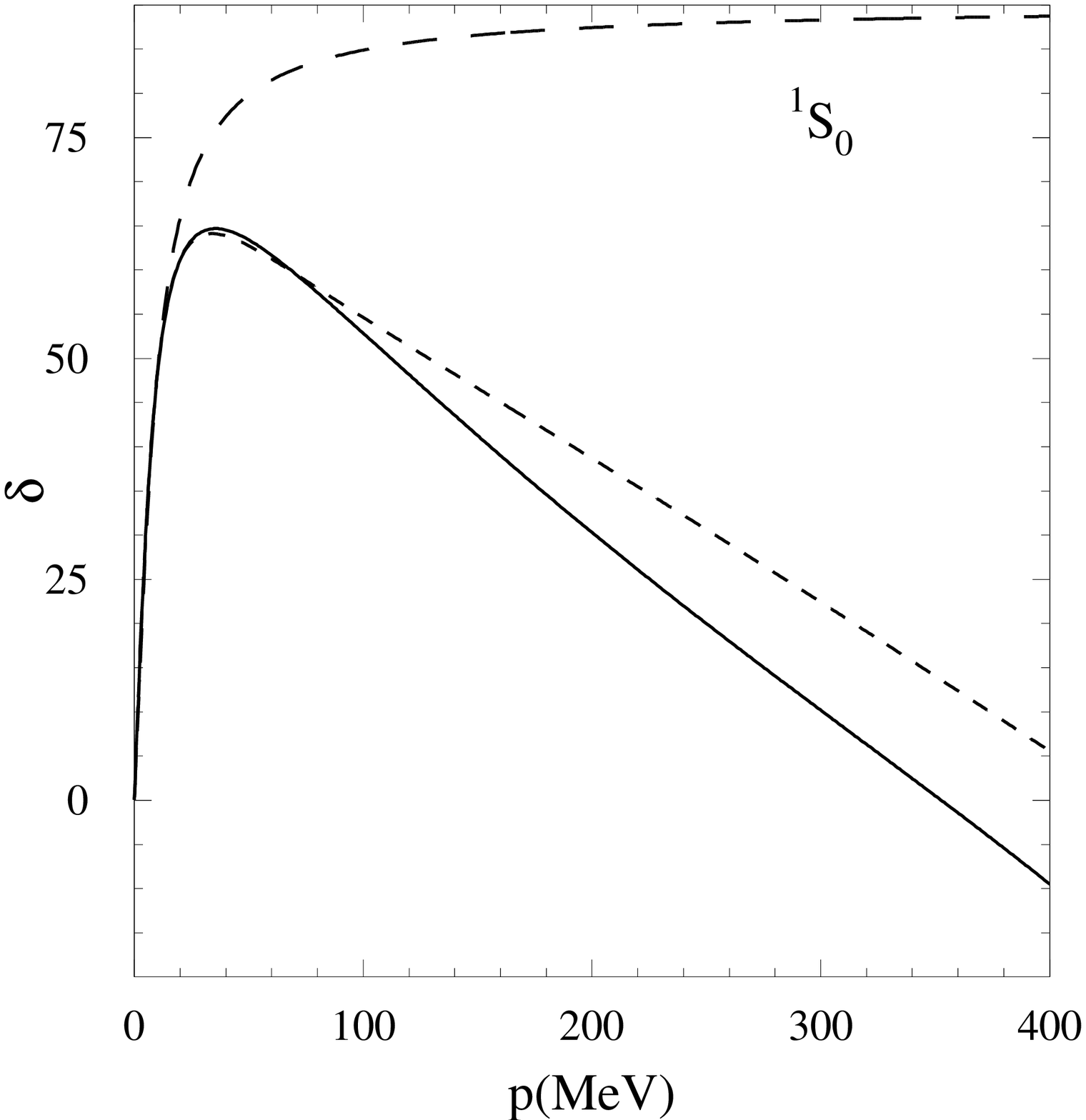}
        \epsfxsize=8truecm \epsfbox{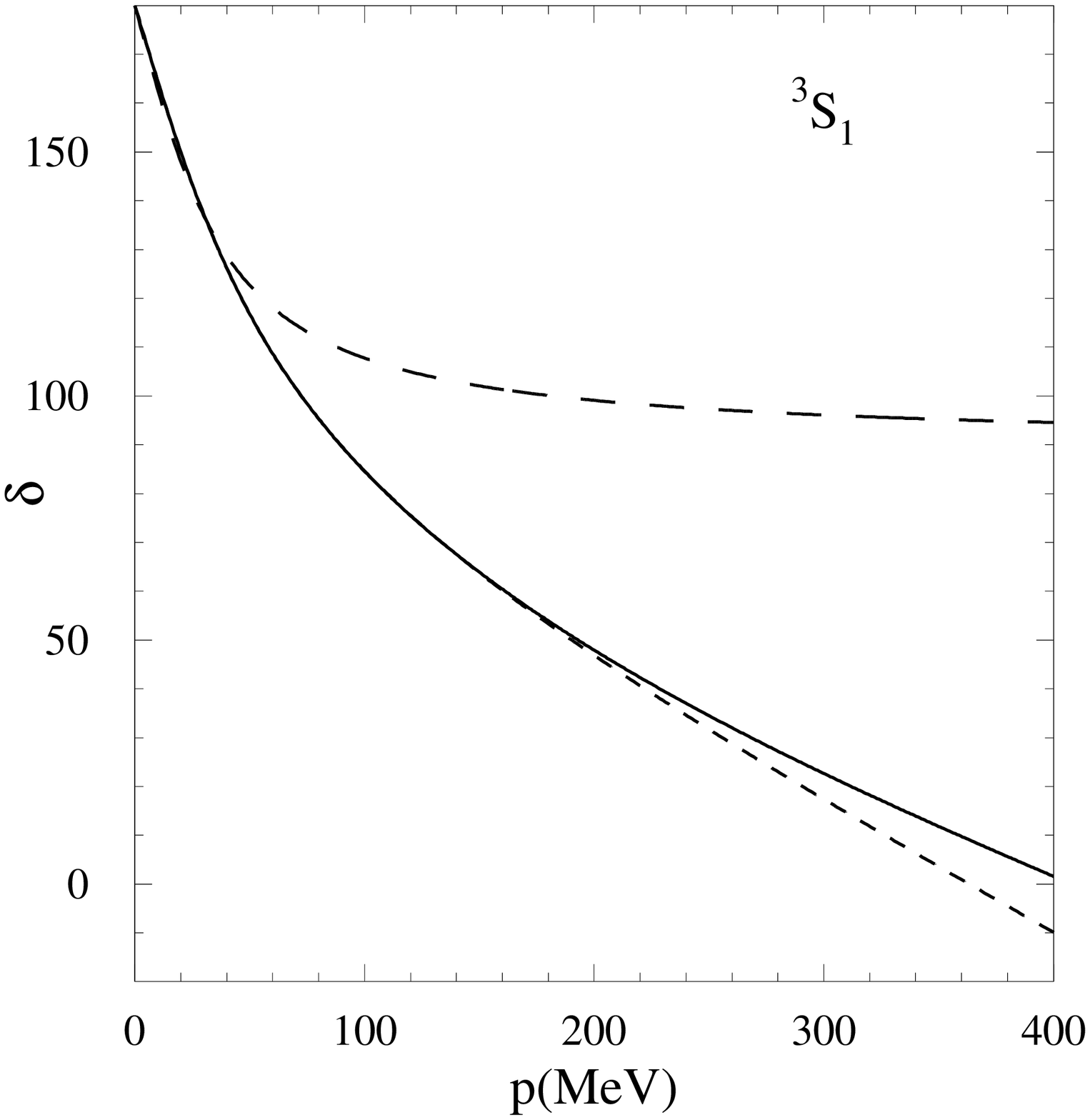} }
\vspace{.3cm}
 {\tighten  
\caption[1]{Fit to the phase shift data weighted toward low momentum.  The solid
line is the Nijmegen fit to the data \cite{Nij}, the long dashed line is the order $1/Q$
result, and the short dashed line is the order $Q^0$ result. } \label{fig_fits} }
\end{figure}

Fits are performed to the Nijmegen phase shift \cite{Nij} data between $7$ and
$100\,{\rm MeV}$ for both the $^1S_0$ and $^3S_1$ channels.  The phase shifts
have an expansion of the form $\delta = \delta^{(0)} + \delta^{(1)} + {\cal
O}(Q^2/\Lambda^2)$, where \cite{ksw}
\begin{eqnarray}  \label{deltas}
  \delta^{(0)} &=& -\frac{i}{2} \ln{\bigg[ 1 + i\frac{pM}{2\pi} A^{(-1)}
        \bigg]} \,,\qquad
   \delta^{(1)} = \frac{pM}{4\pi} \frac{ A^{(0)} }{ 1 + i\frac{pM}{2\pi}  A^{(-1)} }\,.
\end{eqnarray}
The fits were weighted towards low momentum since the theoretical error is 
smallest there.  The results are shown in Fig.~\ref{fig_fits}.  We chose to fit using 
the amplitudes in the OS scheme, but an equally good fit is obtained in PDS.  
The parameters $C_0(\mu_R)$, $C_2(\mu_R)$, and $D_2(\mu_R)$ were extracted
for different values of $\mu_R$ in both schemes, and the conditions in
Eq.~(\ref{Mm2PDS}) and (\ref{Mm2OS}) were found to be well satisfied.  For
instance, taking $\mu_R=m_\pi=137\,{\rm MeV}$ in the OS scheme we find
\begin{eqnarray}  \label{fitCs}
  {}^1\!S_0  \qquad  C_0(m_\pi) &=& -3.54\:{\rm fm^2} \,, \qquad 
      C_2(m_\pi) = 3.00 \:{\rm fm^4} \,, \qquad  D_2(m_\pi) = 0.377 \:{\rm fm^4} \,, \\ 
  {}^3\!S_1  \qquad  C_0(m_\pi) &=& -5.81\,{\rm fm^2} \,, \qquad 
      C_2(m_\pi) = 10.16 \:{\rm fm^4} \,, \qquad  D_2(m_\pi) = -4.128 \:{\rm fm^4} \,.\nn
\end{eqnarray} 
The fit gives scattering lengths $a(^1S_0)=-23.3\,{\rm fm}$, and
$a(^3S_1)=5.41\,{\rm fm}$.  Plugging the parameters extracted from the fit into the
right hand side of Eq.~(\ref{Mm2OS}) gives $\sim 10^{-2}$ for both channels.  In
the OS scheme, the renormalization group equations are obeyed to a few percent 
accuracy since the amplitude is explicitly $\mu_R$ independent.  

It is important to realize that the fits do not unambiguously determine the values of
$C_0(\mu_R)$ and $D_2(\mu_R)$.  The coefficient of the four nucleon operator with
no derivatives is $C_0^{\,bare} + m_\pi^2 D_2^{\,bare}$.  Once we switch to
renormalized coefficients, it is not clear how to divide the couplings into a
nonperturbative and perturbative piece. In Ref.\cite{ksw}, $C_0(\mu)$ is summed to
all orders, while other authors \cite{Gegelia2,sf2} treat both coefficients
non-perturbatively.  In fact, in order to do a chiral expansion, $m_{\pi}^2
D_2(\mu_R)$ should be treated perturbatively. Since $m_{\pi}^2 D_2(\mu_R) \sim
Q^0$, this is consistent with the power counting and the renormalization group
equation.  

On the other hand, there is some freedom in dividing $C_0(\mu_R)$ into
nonperturbative and perturbative pieces: $C_0(\mu_R) = C_0^{np}(\mu_R)+
C_0^p(\mu_R)$, where $C_0^{np}(\mu_R)\sim 1/Q$ and $C_0^p(\mu_R) \sim Q^0$. 
This is simply a reorganization of the perturbative series.  For instance, consider
the following expansion of the amplitude in the theory without pions:
\begin{eqnarray}
 A &=& \frac{4\pi}{M} \left[ \frac{1}{-1/a + \frac{r_0}{2} p^2 +... -ip} \right] =  
   \frac{4\pi}{M} \left[ \frac{1}{-1/a -\Delta + \Delta + \frac{r_0}{2} p^2 +... -ip}
   \right]  \nn \\
 & =& \frac{-4\pi}{M} \left[\frac{1}{1/a+\Delta+ip} +\frac{ \frac{r_0}{2} p^2 +
    \Delta}{(1/a+\Delta+ip)^2} + ... \right] \, , 
\label{newexp}
\end{eqnarray}
where $\Delta \lesssim 1/a$.  The series with $\Delta = 0$ and with $\Delta \neq 0$
will both reproduce effective range theory, but differ in the location of the pole that
appears at each order in the perturbative expansion.   In the $^3S_1$ channel, the
pole of the physical amplitude is at $-ip=\sqrt{M E_d}= 45.7\, {\rm MeV}$, where
$E_d$ is the binding energy of the deuteron\footnote{\tighten In fact, in the
${}^3\!S_1$ channel the fit value of $C_0(m_\pi)$ from Eq.~(\ref{fitCs}) gives
$\gamma=47.3 \,{\rm MeV}$.}.  For comparison, $1/a = 36.3\,{\rm MeV}$ in this
channel. For $\Delta =0$, the pole that appears at each order in the perturbative
expansion will be off by 30\%.   For some calculations, such as processes involving
the deuteron \cite{ksw3}, a better behaved perturbation series is obtained by
choosing $1/a +\Delta = \sqrt{M E_d}$.  If we want to reproduce the expansion in
Eq.~(\ref{newexp}) in the theory without pions then part of $C_0(\mu_R)$ must be
treated perturbatively.

In the theory with pions and a non-vanishing $C_0^p(\mu_R)$, the amplitude can 
be obtained from  Eqs.~(\ref{Amp1}--\ref{Amp4}) by substituting $C_0(\mu_R) 
\to C_0^{np}(\mu_R)$ and 
\begin{eqnarray}
     A^{(-1)} \rightarrow A^{(-1)} - [A^{(-1)}]^2   \frac{C_0^p(\mu_R)}
    {C_0^{np}(\mu_R)^2} \, .
\end{eqnarray}
In the OS scheme, the renormalization group equation makes
$C_0^{p}(\mu_R)/[C_0^{np}(\mu_R)]^2$ equal to a constant.  Therefore,
$C_0^p(\mu_R)$ can be simply absorbed into the definition of $D_2(\mu_R)$. 
Because of this, the value of $D_2(\mu_R)$ extracted from fits to NN scattering
data may differ from the value of the renormalized coupling in the Lagrangian. In
the PDS scheme, the renormalization group equation for $C_0^p(\mu_R)$ is
\begin{eqnarray}
 \mu_R \frac{\partial C_0^p(\mu_R)}{\partial \mu_R} &=& 2\, \frac{M\mu_R}{4\pi} 
 C_0^{np}(\mu_R) \bigg[ C_0^p(\mu_R) + \frac{g^2}{2 f^2} \bigg]  \,,
\end{eqnarray} 
with solution 
\begin{eqnarray}
\frac{C_0^p(\mu_R)}{C_0^{np}(\mu_R)^2} +
\frac{g^2}{2 f^2} \frac1{C_0^{np}(\mu_R)^2} = \mbox{constant} \,.
\end{eqnarray}
Therefore, breaking $C_0(\mu_R)$ into perturbative and nonperturbative pieces
results in a manifestly $\mu_R$ independent amplitude in PDS\footnote{We would
like to thank Mark Wise for pointing this out to us.}.  The constraint in
Eq.~(\ref{Mm2PDS}) is now $\mu_R$ independent.

\begin{figure}[!t]  
  \centerline{\epsfxsize=8.0truecm \epsfbox{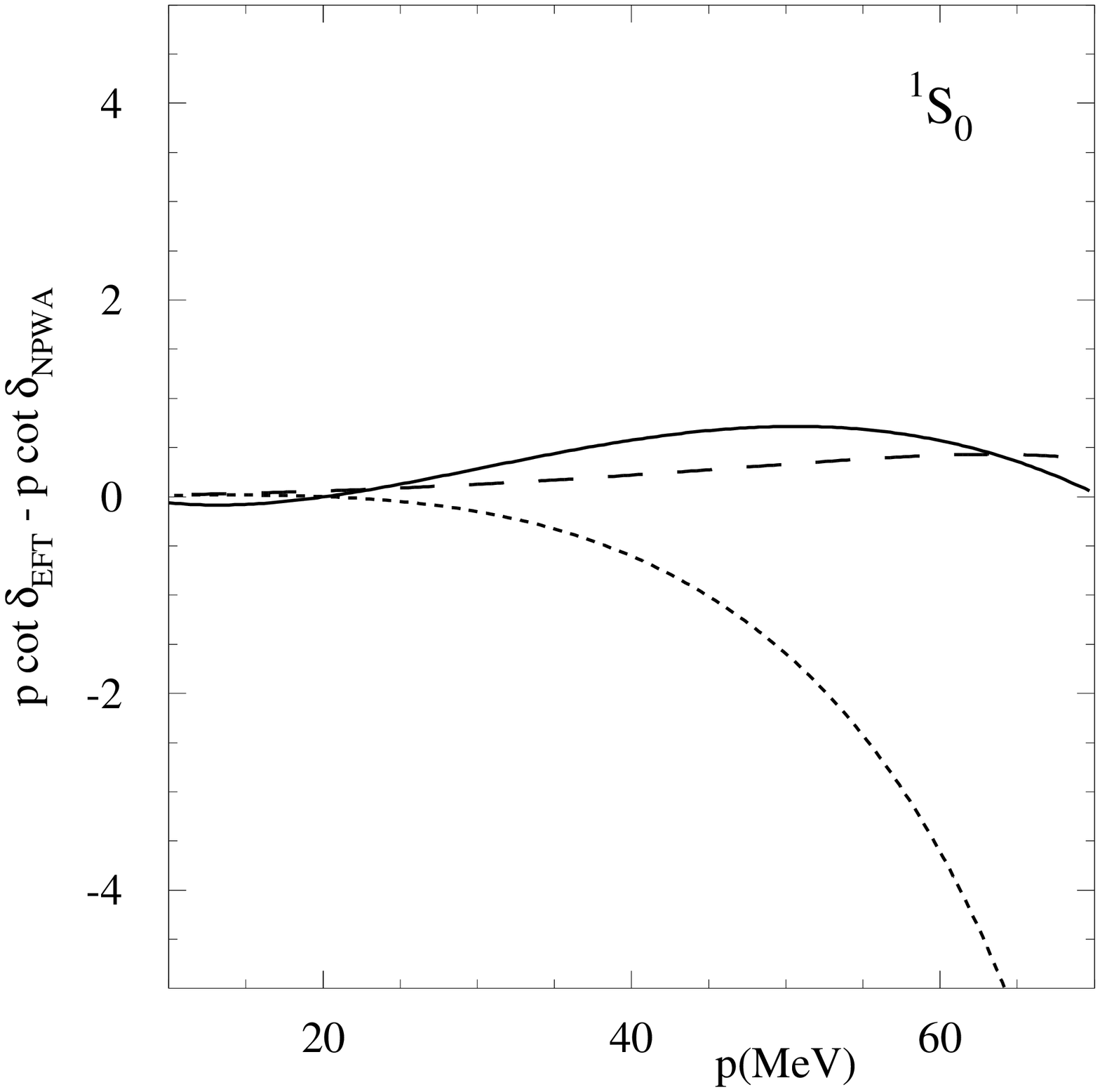}
        \epsfxsize=8truecm \epsfbox{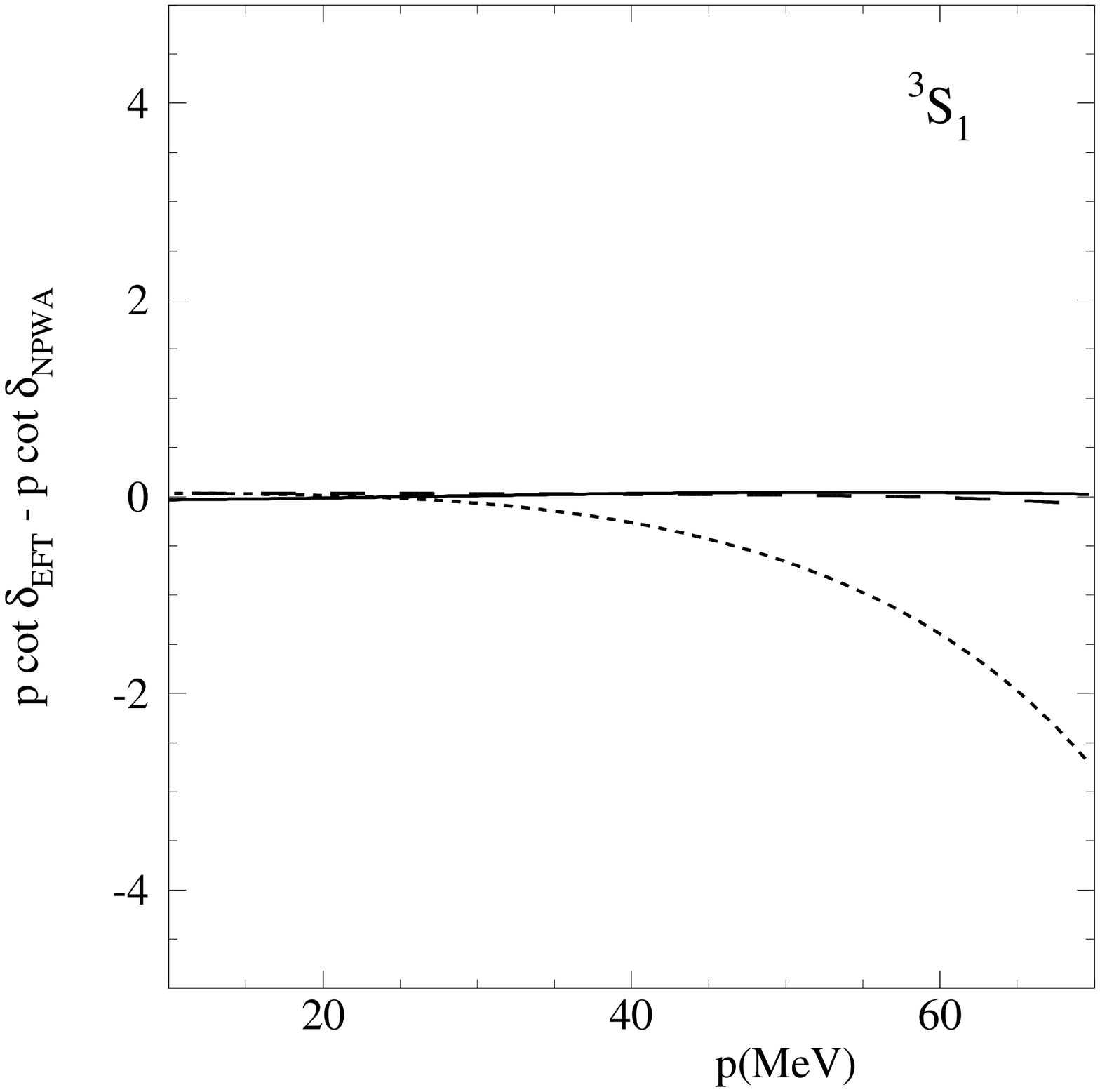} }
 {\tighten  
\caption[1]{The effective field theory  and Nijmegen Partial Wave analysis \cite{Nij}
values of $p\cot\delta$ are compared.  The solid lines use $p\cot(\delta^{(0)}+
\delta^{(1)})$, the dashed lines use Eq.~(\ref{erexpn}) with the 
$v_i$ from Ref.~\cite{Cohen2}, and the dotted lines use the values of $v_i$ 
from the low-energy theorems.  } \label{fig_ch} }
\end{figure}
Integrating out the pion gives low-energy theorems for the coefficients $v_i$ in 
the effective range expansion\cite{Cohen2},
\begin{eqnarray} 
  p\cot{\delta(p)} = -\frac{1}{a} + \frac{r_0}{2}\, p^2 + v_2\, p^4 + v_3\, p^6 +
     v_4\, p^8 + \ldots \,.  \label{erexpn}
\end{eqnarray}
The $v_i$ can be predicted in terms of one parameter, $C_0^{np}(\mu_R)$, which is
fixed in Ref.~\cite{Cohen2} by the condition $4\pi/[M C_0^{np}(\mu_R)]+\mu_R =
1/a$.  Corrections to these predictions are expected to be $30-50\%$ due to higher
order $Q/\Lambda$ terms.  The $v_i$ extracted from the phase shift data
\cite{Cohen2,Nij2} disagree with the low-energy theorems by factors of order $5$. 
In Fig.~\ref{fig_ch}, we see that the agreement of $p\cot(\delta^{(0)} + \delta^{(1)})$
(solid lines)\footnote{\tighten Note that when expanded in $Q$, $p\cot\delta= ip
+4\pi/[MA^{(-1)}] -4\pi A^{(0)}/[M(A^{(-1)})^2]+{\cal O}(Q^3)$, which differs from
$p\cot(\delta^{(0)}+\delta^{(1)})$ by terms of order $Q^3$. The latter expression is
used since the parameters in Eq.~(\ref{fitCs}) were fit using Eq.~(\ref{deltas}).} with
the Nijmegen partial wave analysis is comparable to that of the effective range
expansion with the $v_i$ from the fits in Refs.~\cite{Cohen2,Nij2} (dashed lines). 
Note that our fit is more accurate at low momentum than the global fit in
Ref.~\cite{ksw}.  However, keeping only the first five terms from the low-energy
theorems (dotted lines) gives larger disagreement at $70\,{\rm MeV}$.  This is not
surprising since the pion introduces a cut at $p= i \,m_\pi/2$, so the radius of
convergence of the series expansion of $p \cot{(\delta)}$ in Eq~(\ref{erexpn}) is
$\simeq 70\, {\rm MeV}$.  At $p=70\,{\rm MeV}$, one expects large corrections from
the next term in the series.  However, the fit values of $v_i$ give good agreement
with the data even at $70\,{\rm MeV}$.  It is possible that uncertainty from higher
order terms in the Taylor series has been absorbed into $v_2$, $v_3$, and $v_4$ in
the process of performing the fits.  For this reason, the uncertainty in the values of
$v_i$ that were found from fitting to the data may be considerable.  

To get an idea of the error in $v_2$, we will specialize to the $^3S_1$ channel.  The
Nijmegen phase shift analysis \cite{NijPW} lists two data points for $p < 70\,{\rm
MeV}$: $ p= 21.67 \,{\rm MeV}$ where $\delta^{(^3S_1)}= 147.747 \pm 0.010\,^\circ$,
and $p= 48.45 \,{\rm MeV}$, where $\delta^{(^3S_1)}= 118.178 \pm 0.021\,^\circ$. 
Using $a=5.420 \pm 0.001 \,{\rm fm}$ and $r_0 = 1.753 \pm 0.002\,{\rm fm}$ \cite{Nij2}
in the effective range expansion and fitting to the lowest momentum data point, we
find $v_2 = -0.50\, \pm\, 0.52 \,{\rm fm^3}$, where the error in $a$, $r_0$, and
$p\cot\delta $ have been added in quadrature.  This differs by one sigma from both
the value predicted by the low-energy theorem, $v_2^{\rm thm}=-0.95 \,{\rm
fm^3}$, and the value from the fit, $v_2^{\rm fit} = 0.04 \,{\rm fm^3}$.  Since the
range of the pure nucleon theory is $70\,{\rm MeV}$, there will also be a $\simeq 0.1
\,{\rm fm^3}$ error in this extraction from $v_3$ and higher coefficients.  This error
was estimated by comparing the theoretical expression for $p\cot\delta$ with the
first three terms in its series expansion.  If we instead use the higher momentum
point we find $v_2 = 0.03\, \pm\, 0.04 \,{\rm fm^3}$ with $\simeq 0.5\,{\rm fm^3}$
theoretical uncertainty.  The uncertainty in these values of $v_2$ is too large to
make a definitive test of the low-energy theorems.  

In this paper, we have addressed the issue of $\mu_R$ sensitivity in perturbative
treatments of the pion. Amplitudes are $\mu_R$ independent in the OS scheme,
and in the PDS scheme, if part of $C_0(\mu_R)$ is treated perturbatively.  Fits to
NN scattering data were done which agree well at low momentum.  Errors at high
momentum are consistent with uncertainty from higher order terms if the range of
the effective field theory is $\gtrsim 300\,{\rm MeV}$.  We conclude that there is no
obstruction to using perturbative pions for momenta $p>m_\pi$.  In a future
publication\cite{ms1}, we will describe the renormalization procedure and OS
scheme in greater detail.  The range of the theory with perturbative pions will also
be investigated.

We would like to thank Mark Wise for many useful conversations.  We also would
like to thank H. Davioudiasl, S. Fleming, U. van Kolck, Z. Ligeti, S. Ouellette and K. 
Scaldefferri, for their comments.  T.M would like to acknowledge the hospitality of
the Department of Physics at the University of Toronto, where part of this work
was completed.  This work was supported in part by the Department of Energy
under grant number DE-FG03-92-ER 40701.

{\tighten

} 

\end{document}